\documentclass[aps,pra,twocolumn,amssymb,amsfonts]{revtex4-1}
\usepackage{multirow} 
\usepackage{bm}
\usepackage{dcolumn}
\usepackage{bm}
\usepackage{mathbbol}
\pdfoutput=1
\usepackage{graphicx}
\usepackage{amsmath,amsthm}
\usepackage{times}
\usepackage{color}

\begin{document}
\title{Extended nonergodic states in disordered many-body quantum systems}

\author{E. J. Torres-Herrera$^1$ and Lea F. Santos$^2$}
\affiliation{$^1$Instituto de F{\'i}sica, Universidad Aut\'onoma de Puebla, Apt. Postal J-48, Puebla, Puebla, 72570, Mexico}
\affiliation{$^2$Department of Physics, Yeshiva University, New York, New York 10016, USA}

\date{\today}

\begin{abstract}
This work supports the existence of extended nonergodic states in the intermediate region between the chaotic (thermal) and the many-body localized phases. These states are identified through an extensive analysis of static and dynamical properties of a finite one-dimensional system with onsite random disorder. The long-time dynamics is particularly sensitive to changes in the spectrum and in the structures of the eigenstates. The study of the evolution of the survival probability, Shannon information entropy, and von Neumann entanglement entropy enables the distinction between the chaotic and the intermediate region. 
\end{abstract}

\maketitle

\section{Introduction}

The Anderson localization in real space of a particle in a disordered medium~\cite{Anderson1958}  is theoretically well understood~\cite{Lee1985} and experimentally confirmed in various settings~\cite{Kramer1993,Billy2008,Roati2008,Modugno2010,McGehee2013}. In systems with many interacting particles, despite the consensus that many-body localization should still take place in the limit of strong disorder, the details of the metal-insulator transition are richer than in noninteracting systems and not yet completely understood.

The seeds for the analysis of many-body localization were sown in the 80's \cite{Fleishman1980,Altshuler1988,Giamarchi1988} and 90's \cite{Dorokhov1990,Shepelyansky1994,Imry1995,Altshuler1997,Guhr1998}. The general expectation was that interaction might hamper localization, but not completely prevent it. The study of two interacting particles in a one-dimensional (1D) disordered chain, for instance, showed that the localization length was larger than in a one-particle system~\cite{Shepelyansky1994}. A number of theoretical studies followed in the 2000's \cite{Georgeot2000,Berman2001,Santos2004,SantosEscobar2004,Santos2005loc,Gornyi2005,Basko2006,Oganesyan2007,Znidaric2008,Brown2008,Dukesz2009}, but the current decade has witnessed a burst of interest in the subject, with not only theoretical~\cite{Pal2010,Bardarson2012,DeLuca2013,Imbrie2016a,Imbrie2016b,Huse2014,Kjall2014,Lev2014,Lev2015,GroverARXIV,Vosk2015,Chandran2015,Luitz2015,Luitz2016,Ros2015,Torres2015,Torres2016BJP,Yang2015,Altman2015,Nandkishore2015,Serbyn2013,Serbyn2014,Serbyn2015,Serbyn2016,Singh2016,Agarwal2015,Gopalakrishnan2016,Monthus2016,Xiaopeng2015,Xiaopeng2016,Devakul2015,Goold2015,Santos2016,Bertrand2016,Gogolin2016,Cohen2016,Barisic2016,Khemani2016,EnssARXIV}, but also experimental works~\cite{Schreiber2015,Kondov2015,Bordia2016,Smith2015}.

Among the several questions that have been raised, the existence or not of an intermediate phase between the chaotic and the many-body localized phase~\cite{Altshuler1997,GroverARXIV,Xiaopeng2015,Xiaopeng2016,DeLuca2014,Goold2015,Kravtsov2015,Znidaric2016} is the one that mainly motivates the present work. Our numerical results for a finite 1D spin-1/2 system with nearest-neighbor interaction and onsite random magnetic fields suggest a positive answer. A key aspect of our approach is the shift of the emphasis from spectral properties to dynamical properties of the system. This establishes a strong connection between our studies and current experiments where dynamics is typically investigated. We show that detailed information about the spectrum, eigenstates, and initial state can be obtained from analyzing the system's time evolution.

As the strength of the disorder increases from zero, where the system is integrable, the following regions are covered, before localization is finally reached: (i) transition from the integrable to the chaotic domain, (ii) chaotic regime, and (iii) intermediate region between the chaotic limit and the many-body localized phase. Chaotic states prevail in the chaotic domain, while in the intermediate region before localization, the eigenstates are delocalized (extended) in the configuration space but nonchaotic (nonergodic).

Our definition of a chaotic state is as follows. Given an eigenstate $|\psi^{(\alpha)} \rangle = \sum_n  C^{(\alpha)}_n |\phi_n \rangle$, the inverse participation ratio
\begin{equation}
IPR^{(\alpha)} = \sum_{n=1}^{N} |C^{(\alpha)}_n |^4 
\label{Eq:PR}
\end{equation}
measures how much delocalized the eigenstate is in the basis vectors $|\phi_n \rangle$, with $N$ being the dimension of the Hamiltonian matrix. The eigenstate is chaotic if it samples most of the Hilbert space without any preference, which gives $IPR^{(\alpha)} \propto N^{-1}$. The paradigmatic example of chaotic states are the eigenstates of full random matrices (FRM), which are (pseudo-)random vectors and therefore maximally delocalized. When these matrices are the real and symmetric ones from Gaussian orthogonal ensembles (GOE) \cite{Wigner1958,MehtaBook,Guhr1998}, the components of their eigenstates are real uncorrelated random numbers from a Gaussian distribution satisfying the normalization condition, so $IPR^{FRM} \simeq 3 N^{-1}$ \cite{ZelevinskyRep1996,Torres2016}. 

In FRM, the notion of basis is not well defined, but in realistic systems, they certainly affect the results for $IPR$. In this case, the choice of basis is done according to the question we are addressing. In studies of spatial localization, the focus is on the basis that represents the configuration space.  Another important aspect is the energy dependence of $IPR$. In realistic finite chaotic systems with two-body interactions, since the density of states is Gaussian~\cite{Brody1981}, it is away from the edges of the spectrum that we expect to find eigenstates that are close to random vectors and have $IPR^{(\alpha)} \propto N^{-1}$, while near the borders they tend to be more localized. Strict ergodicity in the sense of full random matrices does not hold in these systems.

A delocalized but nonchaotic eigenstate has $IPR^{(\alpha)} \propto N^{-D_2}$ with $D_2<1$. For a localized eigenstate, $IPR^{(\alpha)} \propto {\cal O} (1)$. Our focus is on the middle of the spectrum of regions (ii) and (iii), where we find respectively:
\begin{eqnarray}
&\text{Chaotic states:}  & IPR \propto N^{-1} , \\
&\text{Delocalized nonchaotic states:} \hspace{0.1 cm} & IPR \propto\!\! N^{-D_2}, \hspace{0.1 cm}  D_2<1  .
\label{Eq:DelocChaos} 
\end{eqnarray}

The three different regions between the integrable point and the localized phase are distinguished through the analysis of static and dynamical properties. We do not aim at locating exactly the point that separates one region from the other, but discuss the best quantities to identify and characterize each one. System sizes larger than the ones treated here are necessary to delineate the borders and precisely determine how they depend on parameters of the spin-1/2 systems, such as anisotropy, and on the energy of the states. 

Our study of the statistic properties involves level statistics, structure of the eigenstates and structure of the initial states. They are used for comparisons with our results for the dynamics. The latter is the actual center of our attention. 

We analyze the spectrum using nearest-neighbor level spacing distributions and level number variance. The structures of the eigenstates are investigated with the inverse participation ratio, the Shannon information entropy, and the von Neumann entanglement entropy. The distribution of the components of the initial state written in the energy eigenbasis are contrasted with the Porter-Thomas distribution~\cite{Brody1981}.   In regions (i) and (iii),  the level statistics is intermediate~\cite{Avishai2002,Santos2004,SantosEscobar2004} between that of uncorrelated and repelling eigenvalues. The results indicate that region (i) may disappear in the thermodynamic limit. In region (iii), the eigenstates are nonchaotic but delocalized, and there is no agreement with the Porter-Thomas distribution. This region appears to shrink with system size and may turn into a critical point in the thermodynamic limit.
 
For the dynamics, we investigate the time evolution of both the Shannon and entanglement entropies and of the survival probability. The latter is defined as
\begin{equation}
W_{n_0} (t) = \left| \langle \Psi(0) | e^{-i H t} | \Psi(0) \rangle \right|^2, 
\label{Eq:SP}
\end{equation}
where $|\Psi(0) \rangle$ is the initial state, $H$ is the Hamiltonian describing the system, and $\hbar =1$. The decay of the survival probability at long times is necessarily power law, $W_{n_0} (t) \propto t^{-\gamma}$, even in the chaotic regime~\cite{Tavora2016,TavoraARXIV}. The value of the power-law exponent $\gamma$ is very sensitive to the properties of both the initial state and the Hamiltonian. In region (ii), where there is level repulsion following the Wigner-Dyson distribution and the eigenstates are chaotic, we have that $1\leq \gamma \leq 2$. In region (iii), $\gamma<1$ and it coincides with $D_2$ from Eq.~(\ref{Eq:DelocChaos})\cite{Torres2015,Torres2016BJP}. 

The presence of level repulsion is manifested also in the so-called correlation hole~\cite{Pechukas1984,Leviandier1986,Delon1991,Alhassid1992}. This corresponds to a time interval where  $W_{n_0} (t)$ drops below its infinite time average. It happens before saturation takes place and is very evident in region (ii). In region (iii), the correlation hole becomes less deep and fades aways at the approach of the localized phase.

The dynamical behavior of both entropies is very similar. In the chaotic region, they grow linearly in time and quickly saturate, while in region (iii), the linear increase is followed by a logarithmic behavior, before saturation. Since the results for both entropies are comparable, either one can be used in the analysis. The advantage of the Shannon entropy is to be computationally less expensive than the entanglement entropy, because contrary to the latter, it does not require performing a partial trace over degrees of freedom.

\begin{figure}[t!]
\includegraphics[width=7.8cm]{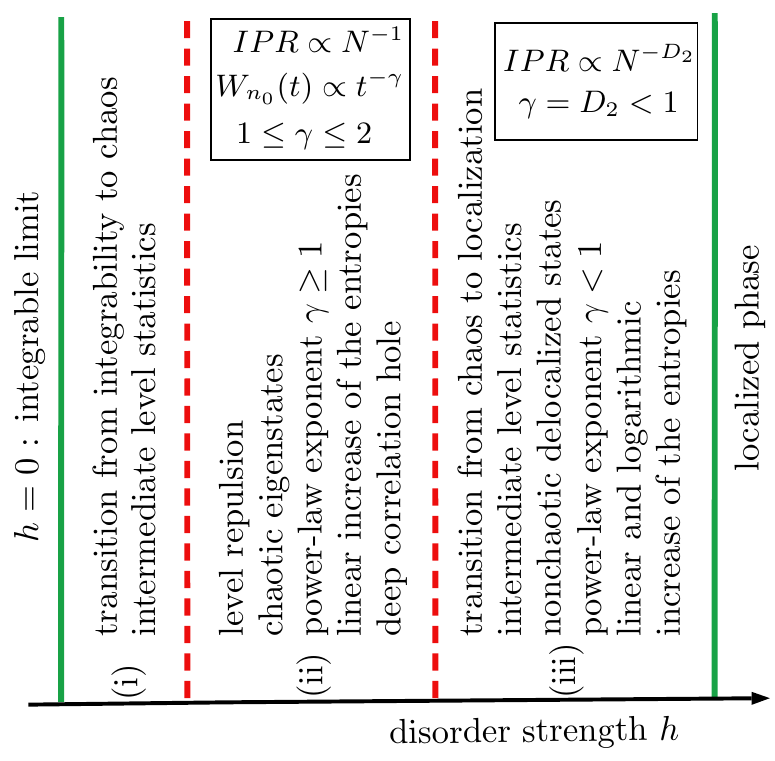}
\caption{\label{Fig:sketch}
The main features of the regions (i), (ii), and (iii) traversed by the finite 1D spin-1/2 model (\ref{ham}) as the strength $h$ of the onsite disorder increases from zero.}
\end{figure}

The main characteristics of regions (i), (ii), and (iii) are summarized in Fig.~\ref{Fig:sketch}.

\section{Model and Static Properties}

The 1D disordered spin-1/2 system that we consider has two-body nearest-neighbor couplings, $L$ sites, and periodic boundary conditions. The Hamiltonian is given by
\begin{equation}
H = \sum_{k=1}^L \left[ h_k  S_k^z  + J \left(
S_k^x S_{k+1}^x + S_k^y S_{k+1}^y +S_k^z S_{k+1}^z \right) \right] ,
\label{ham}
\end{equation}
where $S^{x,y,z}_k $ are the spin operators of each site $k$. We set  $J=1$. The disorder corresponds to random static magnetic fields of amplitude $h_k$, where $h_k$ are random numbers from a uniform distribution $[-h,h]$ and $h$ is the disorder strength. The total spin in the $z$-direction, ${\cal S}^z=\sum_kS_k^z$, is conserved. We study the largest subspace, ${\cal S}^z=0$, which has dimension $N=L!/(L/2)!^2$. 

\subsection{Level statistics}

The observation of intermediate level statistics in the vicinity of the metal-insulator transition dates back to~\cite{Izrailev1989,Shklovskii1993}. In the particular case of many-body quantum systems described by 1D disordered spin-1/2 models, level statistics was studied in~\cite{Avishai2002,Santos2004,SantosEscobar2004}. More recent works include the detailed analyses developed in~\cite{Serbyn2016,Bertrand2016}. Here, the brief discussion of spectral statistics is used for comparison with our results for delocalization measures and dynamics. In addition to the frequently investigated nearest-neighbor level spacing distribution, we consider also the level number variance.

Level repulsion is a main signature of quantum chaos. After unfolding the spectrum~\cite{Guhr1998,Gubin2012} and separating the rescaled eigenvalues by symmetry sectors~\cite{Gubin2012,Santos2009JMP}, level repulsion can be detected, for example, by computing the nearest-neighbor level spacing distribution $P(s)$, where $s$ is the spacing between two neighboring levels. For FRM from GOE or any chaotic real and symmetric Hamiltonian matrix, $P(s)$ is the Wigner-Dyson distribution of shape~${P_{WD}(s)= \frac{\pi }{2} s \exp \left(- \frac{\pi }{4} s^2 \right)}$. In systems where the levels are not prohibited from crossing and where the number of degeneracies are not excessive~\cite{Zangara2013}, the typical distribution is Poissonian, $P_{P}(s)=\exp(-s)$.
To quantify the proximity to the Wigner-Dyson distribution, we use the chaos indicator~\cite{Jacquod1997}
\begin{equation}
\eta = \frac{\int_0^{s_0} [P(s) - P_{WD}(s)] ds}{\int_0^{s_0} [P_P(s) - P_{WD}(s)] ds}, 
\end{equation}
where $s_0$ is the first intersection point of $P_P(s)$ and $P_{WD}(s)$. For a Poisson distribution, $\eta \rightarrow 1$, and for the Wigner-Dyson, $\eta \rightarrow 0$. 

\begin{figure}[t!]
\hspace{0.3cm}
\includegraphics[width=8.1cm]{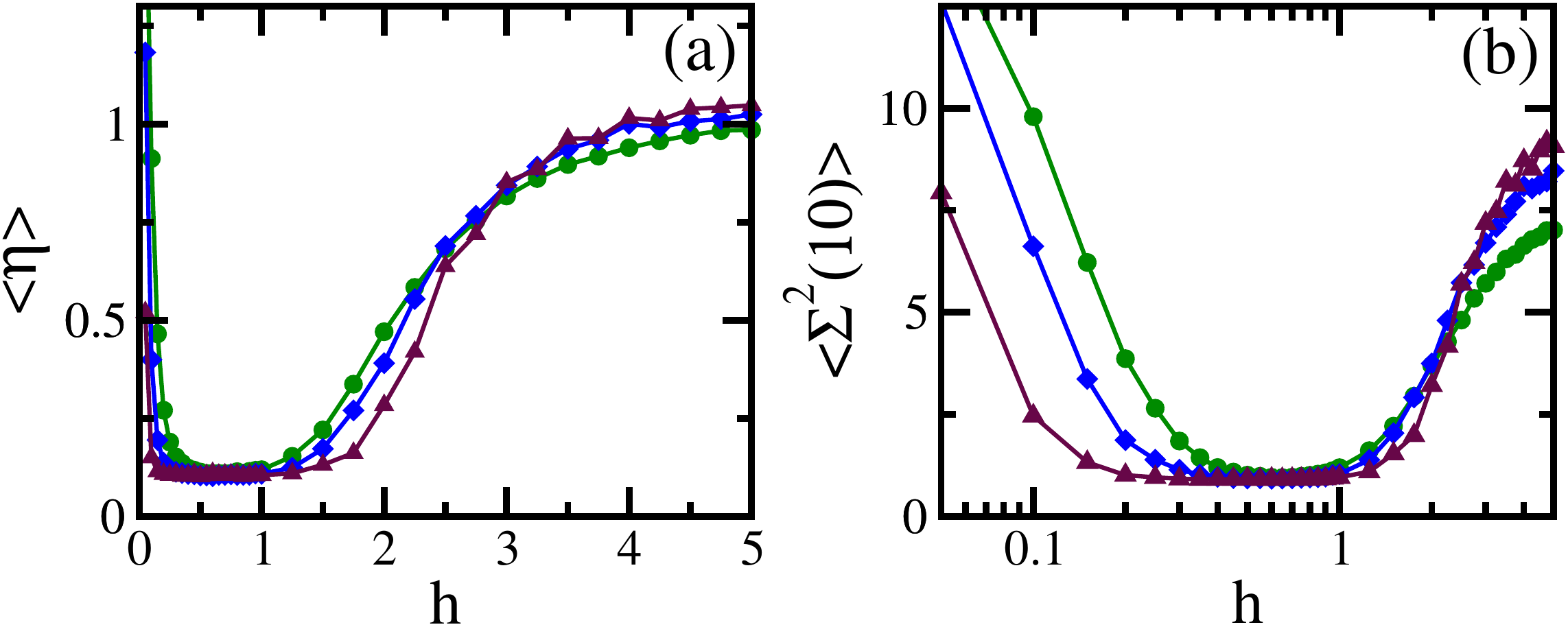}
\caption{\label{Fig:statistics}
Chaos indicator $\langle \eta \rangle $  (a) and level number variance for $\ell=10$ (b) vs. the disorder strength $h$. System sizes $L=12$ (circles), $L=14$ (diamonds), and $L=16$ (triangles). The averages are done over 1082 ($L=12$), 291 ($L=14$), and 77 ($L=16$) disorder realizations. [See explanation for these specific numbers of disorder realizations in Sec. 2.2.1.]}
\end{figure}

In Fig.~\ref{Fig:statistics} (a), we show $\eta$ averaged over several disorder realizations. The average is denoted with $\langle . \rangle $. As $h$ increases from zero, the level spacing distribution for $H$ (\ref{ham}) first transitions abruptly from Poisson ($\eta \sim 1$) to Wigner-Dyson (small $\eta$) in region (i).  It is Wigner-Dyson in region (ii), where $\eta$ plateaus to a small value. It then transitions from Wigner-Dyson back to Poisson in region (iii).  The boundary of region (iii) after which localization emerges occurs approximately where the curves for different system sizes cross, seen in the figure for $ h \in [2.5,3.5]$. Beyond this point, the larger the system is the closer the distribution is to $P_P(s)$.

The level spacing distribution provides information about the short-range fluctuations of the spectrum. Information about long-range correlations, which better captures how rigid the spectrum is, can be obtained with quantities such as the level number variance $\Sigma^2 (\ell)  \equiv (\overline{N(\ell ,\epsilon))^2}  - \left( \overline{N(\ell ,\epsilon)}\right)^2$,
where $N(\ell ,\epsilon)$ is the number of unfolded energy levels $\epsilon$ in the interval $[\epsilon,\epsilon+\ell ]$
and the bar denotes here the average over different initial values of $\epsilon$ \cite{Guhr1998}. For FRM from GOE, $\Sigma_{FRM}^2 (\ell) = \frac{2}{\pi^2} \left( \ln(2 \pi \ell) + \gamma_e + 1 -\frac{\pi^2}{8} \right)$, where $\gamma_e = 0.5772\ldots $ is the Euler's constant. Uncorrelated levels lead to $\Sigma^2 (\ell) =\ell$. A careful study of the level number variance in disordered interacting systems and how it connects with the existence of the Thouless energy was done in~\cite{Bertrand2016}. 

In Fig.~\ref{Fig:statistics} (b), we show how $\langle \Sigma^2 (\ell=10) \rangle $ varies with the disorder strength. The results are very similar to those in Fig.~\ref{Fig:statistics} (a). To draw attention to region (i), we show $h$ in a logarithmic scale. In this region, the disorder strength for which $\langle \eta \rangle$ and $\langle \Sigma^2 (10) \rangle$ reach the smallest values decreases as the system size $L$ increases. This suggests that in the thermodynamic limit, region (i) may disappear and an infinitesimally small $h$ should take the system into the chaotic domain~\cite{Santos2010PRE,Torres2014PRE}. The chaotic region (ii) stretches with $L$ not only in the direction of smaller $h$'s, but also for larger disorder strengths. Whether region (iii) will also disappear or persist in the thermodynamic limit, possibly as a critical point, is an open question.

We stress that the values of $\eta$ and especially of $\Sigma^2$ are very sensitive to the unfolding procedure used~\cite{Bertrand2016,Gomez2002}. The results suffer also from finite size effects and, in the case of $h\rightarrow 0$, from additional symmetries. Hence, the purpose of our illustrations is to support the existence of different regions associated with different values of $h$, but not to exactly delineate their boundaries.

\subsection{Delocalization measures}

The eigenstates provide much information about the system. One way to study their structures is with the moments $IPR_q^{(\alpha)} = \sum_{n=1}^{N} |C^{(\alpha)}_n |^{2q}$.  One finds, for example, that at the Anderson transition critical point, the scaling analysis of $IPR_q^{ (\alpha) }$ with the dimension $N$ (that is, $IPR_q^{ (\alpha) }  \sim N^{-(q-1) D_q}$) leads to a nonlinear dependence of the generalized fractal dimension $D_q$ on $q$. This indicates that the eigenstates become multifractal \cite{Evers2008}. Here, we focus on $IPR_2^{ (\alpha) } $ and drop the subscript $q=2$. This corresponds to the inverse participation ratio defined in Eq.~(\ref{Eq:PR}).

Fractal dimensions can also be obtained from ${D_q = \lim _{N \rightarrow \infty} S_q^{ (\alpha) }/\ln N}$, where  ${S_q^{ (\alpha) }= - (1-q)^{-1} \ln \left( \sum_{n=1}^N |C_n^{(\alpha)}|^{2q} \right)}$ is the R\'enyi entropy~\cite{Atas2014}. When $q\rightarrow 1$, the R\'enyi entropy coincides with the Shannon information entropy, 
\begin{equation}
S_{Sh}^{ (\alpha) }=- \sum_{n=1}^N |C_n^{(\alpha)}|^2 \ln |C_n^{(\alpha)}|^2.
\end{equation}
Like the inverse participation ratio, $S_{Sh}^{ (\alpha) }$ is extensively used to measure the level of delocalization of quantum states~\cite{ZelevinskyRep1996}. In this work, we compute $D_2$ from $IPR^{ (\alpha) }$ an $D_1$ from $S_{Sh}^{ (\alpha) }$.

\subsubsection{Inverse Participation Ratio}\label{secIPR}

\begin{figure}[t!]
\includegraphics[width=8.cm]{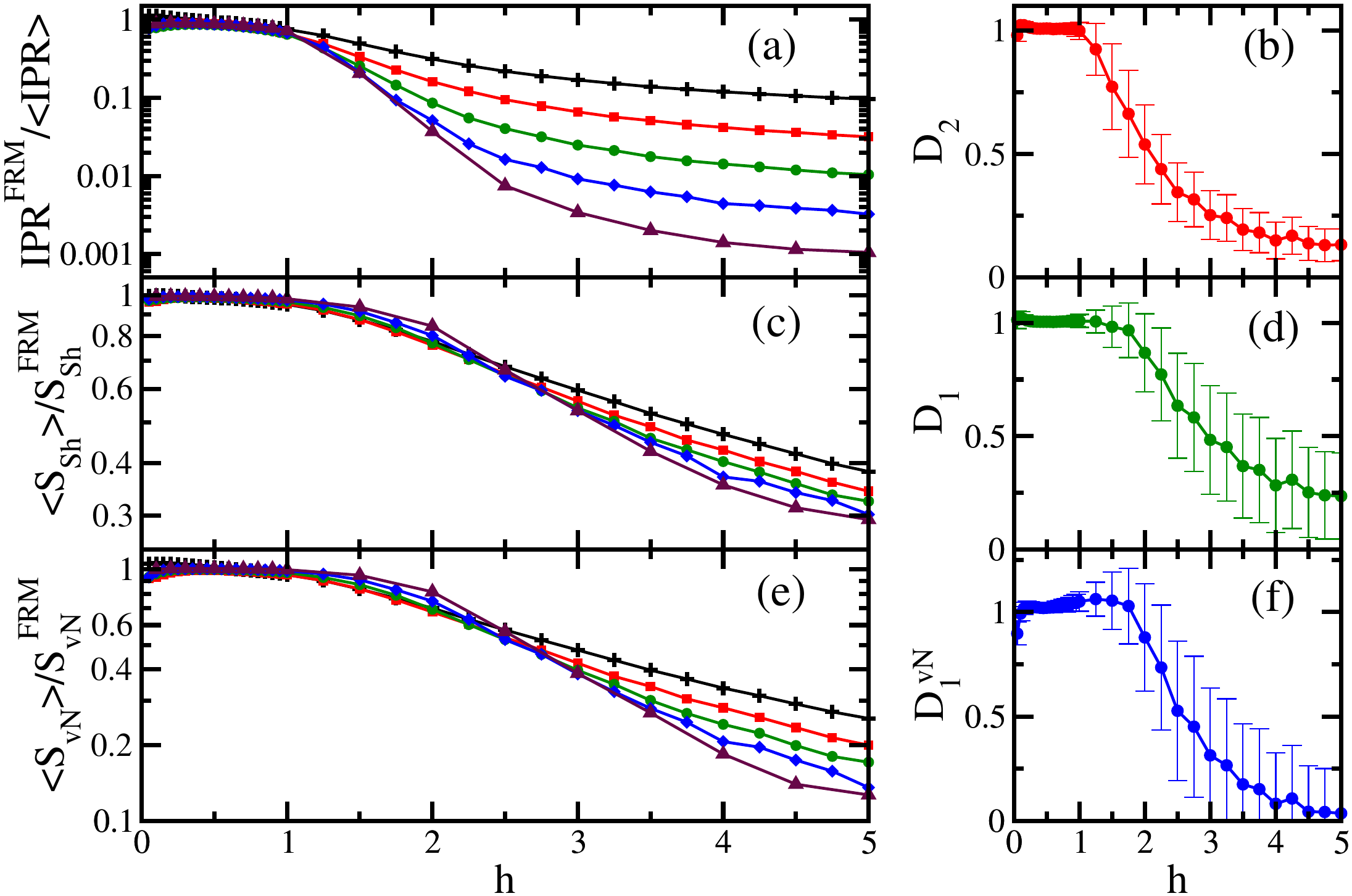}
\caption{\label{Fig:Shannon}
Ratio between the results for FRM with the inverse participation ratio (a), Shannon information entropy (c), and von Neumann entanglement entropy (e), as well as $D_2$ (b), $D_1$ (d), and $D_1^{vN}$ (f) vs. the disorder strength $h$. System sizes $L=8$ (plus), $L=10$ (square), $L=12$ (circle), $L=14$ (diamonds), and $L=16$ (triangles). The averages are done over $\sim 10^5$ data. Error bars in (b), (d), and (e) correspond to standard deviations.}
\end{figure}

Figure~\ref{Fig:Shannon} (a) shows the ratio $IPR^{FRM}/\langle IPR \rangle$, where $\hspace{0.02 cm}$ $IPR^{FRM} = 3/N$. We average $IPR^{(\alpha)}$ over 10\% of the eigenstates that have the closest energies to the middle of the spectrum, where they are more delocalized, and over several disorder realizations, so that the total amount of data is $10^5$. Table~1 gives the specific values for each system size.
\begin{table}[h!]
\begin{center}
  \begin{tabular}{ c | c | c }
    \hline
    System & Number of States  & Number of   \\ 
    Size & from the & Disorder  \\ 
     $L$ & Middle of the Spectrum & Realizations \\ \hline
    8 & 8 & 14 285 \\ \hline
    10 & 26 & 3968 \\ \hline    
    12 & 93 & 1082\\ \hline    
    14 & 344 & 291 \\ \hline    
    16 & 1288 & 77 \\ \hline
  \end{tabular}
  \caption{Number of states, $0.1 N+1$, (they include the eigenstate right at the center of the spectrum and the 5\% that are above and below this energy), and disorder realizations considered for each system size. Approximately $10^5$ data are used for the averages.}
\end{center}
\end{table}

The value of $IPR^{(\alpha)}$ depends, of course, on the basis selected. To study localization in real space, the natural basis corresponds to the states where on each site the spin either points up or down in the $z$ direction, as for example $|\downarrow  \uparrow \downarrow \uparrow \downarrow \uparrow \downarrow \uparrow \ldots \rangle_z$. We refer to these states as site-basis vectors; in quantum information theory they are called computational basis vectors.

The behavior of $IPR^{FRM}/\langle IPR \rangle$ from region (i) to (iii) is nonmonotonic with $h$. This is better seen in a linear plot \cite{Dukesz2009,footDelta} than in the log-plot of Fig.~\ref{Fig:Shannon} (a). The highest level of delocalization occurs in the chaotic region (ii). Notice also that $IPR^{FRM}/\langle IPR \rangle<1$ for all $h$'s, as expected for sparse banded Hamiltonian matrices~\cite{Torres2016}. 

The curves of the ratio $IPR^{FRM}/\langle IPR \rangle$ for different $L$'s cross at a value of $h \in [1,2]$. The fact that after this point the ratio decreases with $L$ implies that $\langle IPR \rangle \propto N^{-D_2}$ (equivalently, $IPR^{FRM}/\langle IPR \rangle \propto 1/N^{1 - D_2}$) with $D_2<1$. The eigenstates are no longer chaotic, although they are still delocalized. This is the beginning of region (iii), while the crossings in Fig.~\ref{Fig:statistics} mark its end. 

The values of the fractal dimension $D_2$ as a function of the disorder strength are shown in Fig.~\ref{Fig:Shannon} (b). $D_2$ is the slope of the plot of $\ln\langle IPR \rangle$ vs $\ln N$. It is obtained from a scaling analysis with $L=8,10,12,14,16$. Some illustrations  can be found in Ref.~\cite{Torres2015}. The values of $D_2$ decrease significantly in region (iii), while in region (i) they are not far from 1. 

Close to the integrable point,  the eigenstates remain delocalized in configuration space, although for the system sizes considered $D_2\lesssim1$. A question that has been debated due to its importance for the subject of thermalization~\cite{Torres2013,Santos2012PRL,Santos2012PRE,He2013,Rigol2014} is whether for $h \rightarrow 0$ the eigenstates right in the middle of the spectrum are chaotic or not. 

\subsubsection{Shannon and Entanglement Entropies}

With Figs.~\ref{Fig:Shannon} (c) and (e), we compare the Shannon information entropy with the von Neumann entanglement entropy, $S_{vN}$. The latter is vastly employed in quantum information theory to quantify the amount of entanglement in a state~\cite{Amico2008,Laflorencie2016}. It is obtained by dividing the system in subsystems A and B and tracing out the degrees of freedom of one of the two, that is,
\begin{equation}
S_{vN}= - \text{Tr} \left( \rho_A \ln \rho_A \right) ,
\end{equation}
where $\rho_A = \text{Tr}_B \rho$ and $\rho$ is the density matrix of the whole system. The dimension of the Hilbert space of the remaining chain of length $L_A$ is $N_A = 2^{L_A}$. We consider $L_A=L/2$.

The values of the entropies are maximum for the random vectors of FRM. In this case, $S_{Sh}^{FRM} \simeq \ln(0.48 N)$ \cite{ZelevinskyRep1996,Torres2016}. For the entanglement entropy, it was shown that  $S_{vN} \simeq \ln(N_A)-1/2$ when $N_A=N_B$ and $N_A N_B = N$ \cite{Page1993}. In analogy with the Shannon entropy, the alternative approximation $S_{vN}^{FRM} \simeq \ln(0.48 N_A)$ was considered in \cite{Torres2016} for $N_A N_B = N_A^2>N$ and is used here.

Figures~\ref{Fig:Shannon} (c) and (e) display, respectively, the ratios $\langle S_{Sh} \rangle /S_{Sh}^{FRM}$ and $\langle S_{vN} \rangle /S_{vN}^{FRM}$ as a function of the disorder strength for different system sizes. The results for both entropies are very similar and the curves intersect approximately at the same point.  This crossing point has been used to detect the transition to localization~\cite{DeLuca2013,footSd}. It occurs at a value of $h$ similar to the one in Fig.~\ref{Fig:statistics} (a). Since the results for both entropies are similar, either one can be used in the studies of many-body localization. The computational advantage of the Shannon entropy is that it does not involve the trace operation.

Figures~\ref{Fig:Shannon} (d) and (f) give $D_1$ and $D_1^{vN}$ as a function of the disorder strength. They are, respectively, the slopes of the plots of $\langle S_{Sh} \rangle$ vs $\ln N$ and $\langle S_{vN} \rangle$ vs $\ln N_A$. The scaling analysis is done with $L=8,10,12,14,16$. The behaviors of $D_1$ and $D_1^{vN}$ are comparable. In the chaotic region (ii), they are 1. Values slightly above 1 are sometimes seen, but they must be caused by finite size effects, the entanglement entropy being more sensitive to them. Both $D_1$ and $D_1^{vN}$ become smaller than 1 for $h$ close to the point where $D_2$ is also $<1$, that is at the beginning of region (iii). 

We notice that in region (iii), the plots for $\langle S_{Sh} \rangle$ vs $\ln N$ and $\langle S_{vN} \rangle$ vs $\ln N_A$ are still linear, yielding the volume-law scaling of the entropies with system size, but since the slopes are smaller than 1, the eigenstates are no longer chaotic. It is in this region that we expect to find the ``two-component'' structure of the entanglement spectrum described in~\cite{Yang2015}.

\section{Dynamical Properties}

To study the dynamics, we consider as initial state a site-basis vector, $|\Psi(0) \rangle \equiv |\phi_{n_0} \rangle = \sum_{\alpha} C_{n_0}^{(\alpha)} |\psi_{\alpha}\rangle $.  Our analysis is performed for the 10\% site-basis vectors $|\phi_{n_0} \rangle$ that have energy $E_{n_0} = \langle \Psi(0) | H |\Psi(0) \rangle $ closest to the middle of the spectrum. The averages are done over these states and over several disorder realizations (see Table~1). The total amount of data for each $h$ is approximately $10^5$. 

We investigate the behavior of the survival probability given by Eq.~(\ref{Eq:SP}). It can also be written as
\begin{equation}
W_{n_0}(t) = \left|\sum_{\alpha} |C^{(\alpha)}_{n_0} |^2 e^{-i E_{\alpha} t}  \right|^2 = \left| \int \rho_{n0} (E) e^{-i E t} dE  \right|^2,
\label{Eq:W0_t}
\end{equation}
where $\rho_{n0} (E) =\sum_{\alpha}  \left| C_{n_0}^{(\alpha)} \right|^2 \delta (E - E_{\alpha }) $ is the energy distribution of $|\Psi(0) \rangle$ weighted by the components  $|C^{(\alpha)}_{n_0} |^2$. This distribution is referred to as the local density of states (LDOS). The survival probability is the absolute square of the Fourier transform of the LDOS. 

The perturbation that takes the system out of equilibrium is very strong. The Hamiltonian initially has $J=0$, so $|\Psi(0) \rangle =|\phi_{n_0} \rangle$, and the coupling strength is abruptly changed to the finite value $J=1$. Since the strength of the perturbation  is strong, the envelope of the LDOS is Gaussian. This mirrors the shape of the density of states of systems with two-body interactions (see Refs.~\cite{Torres2014PRA,Torres2014NJP,Torres2014PRAb} and references therein for more details). The LDOS may, however, be very sparse. This is what happens in region (iii), where the disorder is strong~\cite{Torres2015} and the energy eigenbasis are no longer chaotic.

At very short times, the decay of the survival probability is quadratic, $W_{n_0}(t) \approx 1 - \sigma_{n_0}^2 t^2 $, where $\sigma_{n_0} = \left(\sum_{\alpha} |C^{(\alpha)}_{n_0} |^2 E_{\alpha}^2 - E_{n_0}^2 \right)^{1/2}$ is the width of the LDOS. Subsequently, for intermediate times, the behavior depends on the shape of the LDOS. The decay remains Gaussian when the envelope of the LDOS is also Gaussian or it switches into an exponential decay when $\rho_{n0} (E)$ is Lorentzian~\cite{Torres2014PRA,Torres2014NJP,Torres2014PRAb}.  For long times, however, it is the filling of the energy distribution of the initial state that determines the behavior of $W_{n_0}(t)$ \cite{Tavora2016,TavoraARXIV}. Independently of how fast the initial evolution may be, the long-time decay necessarily slows down and becomes power law~\cite{Tavora2016,TavoraARXIV}, $W_{n_0}(t) \propto t^{-\gamma}$. The long-time evolution is the focus of Secs.~\ref{Sec:strong} and \ref{Sec:weak}.

We also study the evolution of the entanglement entropy and how the Shannon entropy spreads in time over the other site-basis vectors, that is,
\begin{equation}
S_{Sh}(t)= - \sum_{n=1}^{ N} W_n (t) \ln W_n (t) , 
\label{Eq:Sh_t}
\end{equation}
where $W_n(t) = \left| \langle \phi_n | e^{ - iHt} | \Psi(0)  \rangle  \right|^2  $ is the probability for the initial state to be found in state $|\phi_n \rangle$ at time $t$. For both entropies, the evolution at very short times is $\propto t^2$ \cite{Torres2014NJP,Torres2016}. Our interest is in what happens afterwards.

\subsection{Infinite time averages and Porter-Thomas distribution}

\begin{figure}[t!]
\includegraphics[width=8.2cm]{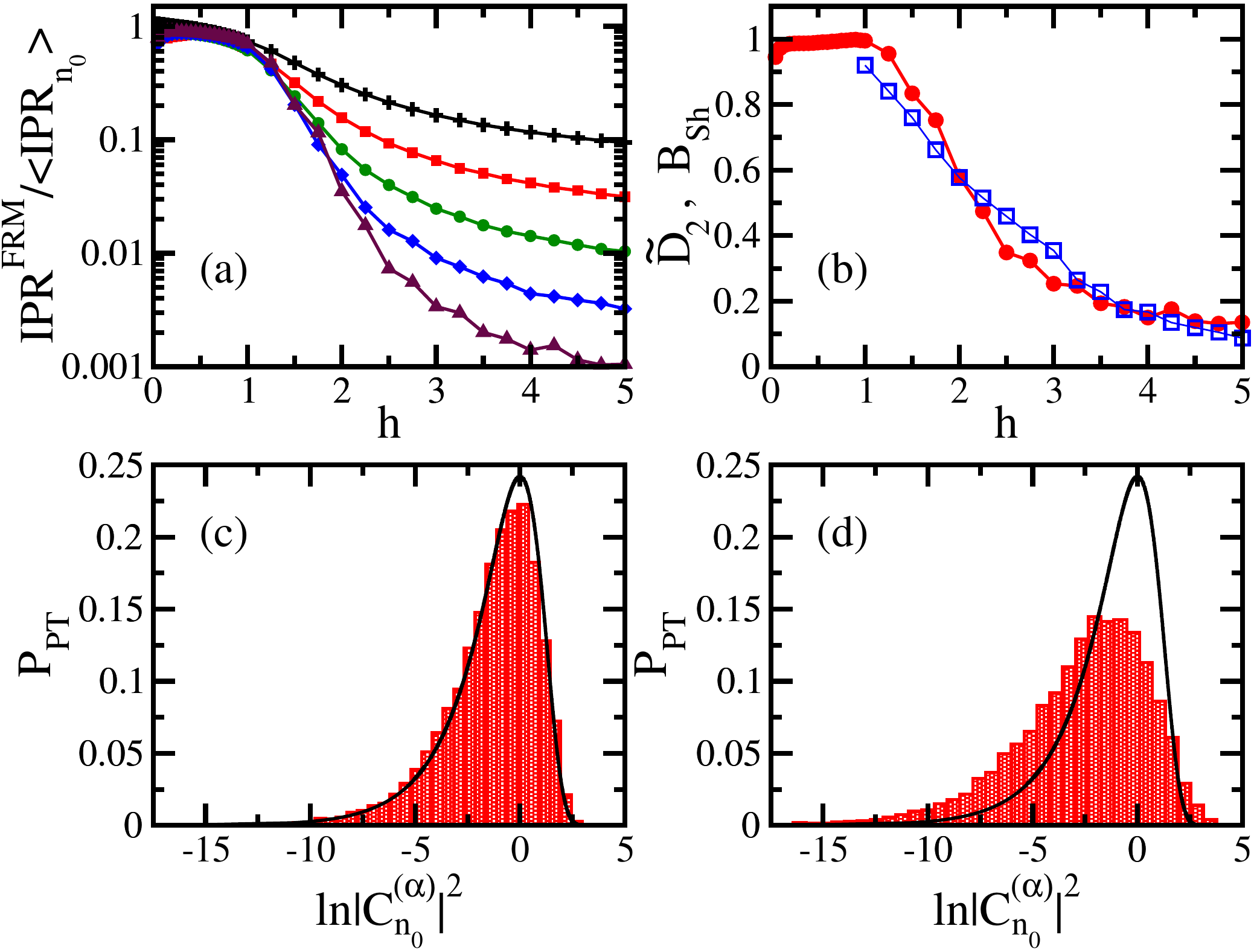}
\caption{\label{Fig:PRini}
$IPR^{FRM}/\langle IPR_{n_0} \rangle$ (a), $\widetilde{D}_2$ (filled circles) and $B_{Sh}$ (open squares) (b) vs. the disorder strength $h$. Porter-Thomas distribution for  $h=0.5$ (c) and $h=2$ (d). In (b), the fitting parameter $B_{Sh}$ from Eq.~(\ref{Eq:B}) is obtained for $L=14$. In (a): System sizes $L=8$ (plus), $L=10$ (square), $L=12$ (circle), $L=14$ (diamonds), and $L=16$ (triangles); they are all used to calculate $\widetilde{D}_2$. The averages are done over $10^5$ data of disorder realizations and initial states. In (c) and (d): one initial state and one disorder realization, $L=16$.}
\end{figure}

We start by analyzing how the structure of the initial state written in the energy eigenbasis depends on $h$. The inverse of the participation ratio of the initial state, $IPR_{n_0}$, coincides with the survival probability after its saturation, that is after it reaches its infinite time average, 
\begin{eqnarray}
\overline{W}_{n_0} &=& \sum_{\alpha} |C^{(\alpha)}_{n_0} |^4  + \overline{\sum_{\alpha \neq \beta} |C^{(\alpha)}_{n_0} |^2 |C^{(\beta)}_{n_0} |^2 e^{i (E_{\beta} -E_{\alpha})t}}  \nonumber \\ 
&\sim & \sum_{\alpha} |C^{(\alpha)}_{n_0} |^4=  IPR_{n_0}
\label{Eq:satF}
\end{eqnarray}
The results for $\langle IPR_{n_0} \rangle $ are shown in Fig.~\ref{Fig:PRini} (a). The scaling analysis gives $\widetilde{D}_2$,  depicted in Fig.~\ref{Fig:PRini} (b). There is great similarity between Figs.~\ref{Fig:PRini} (a), (b) and Figs.~\ref{Fig:Shannon} (a), (b). In fact, $D_2 \sim \widetilde{D}_2$. 

Initial states for which $\widetilde{D}_{2} \sim 1$ are chaotic, which implies the uniform and unbiased sampling of the Hilbert space. In a chaotic initial state, $C_{n_0}^{(\alpha)}$ are approximately zero-centered Gaussian random variables. For a large system, the weights $ |C_{n_0}^{(\alpha)}|^2$ follow the Porter-Thomas distribution~\cite{PorterBook,Brody1981,ZelevinskyRep1996},
\begin{equation}
P_{PT} (|C_{n_0}^{(\alpha)}|^2) = \left(\frac{N}{2\pi |C_{n_0}^{(\alpha)}|^2}\right)^{1/2}\exp{\left(-\frac{N}{2}|C_{n_0}^{(\alpha)}|^2\right)}.
\end{equation}
Notice that the weights appear also in the denominator of the first term. The Porter-Thomas distribution is a result from random matrix theory. It is an additional tool to determine whether the (initial) states are ergodic or not. As the states localize, the components fluctuate more and deviate from the Porter-Thomas. Disagreement with this distribution occurs whenever $D_2,\widetilde{D}_{2} < 1$. 

In Figs.~\ref{Fig:PRini} (c) and (d), we show the distribution of the weights for two different disorder strengths. The logarithmic of  $|C_{n_0}^{(\alpha)}|^2$ is used to better visualize the distribution of the small components, which appear in large amounts. For $h=0.5$, the system is in the chaotic regime ($\widetilde{D}_2\approx 0.99$) and consequently the distribution of the components follows very well the Porter-Thomas distribution. As the disorder strength increases and the system approaches the localized phase, the distribution of the components broadens. In Fig.~\ref{Fig:PRini} (d) we depict the case of $h=2$, where $\widetilde{D}_2\approx0.58$;  the deviation from the Porter-Thomas distribution is evident. 

\subsection{Time evolution under strong disorder}
\label{Sec:strong}

As mentioned above, the long-time decay of the survival probability becomes power law, so the average over initial states and disorder realizations gives $\langle W_{n_0}(t) \rangle \propto t^{-\gamma}$. There are different causes for this algebraic behavior. In region (iii), where the LDOS is sparse, $\widetilde{D}_2<1$  and the eigenstates are correlated. In this case, the power-law exponent $\gamma$ reflects the level of correlations of the eigenstates and  of the components $|C^{(\alpha)}_{n_0} |^2$. We have that $\gamma \sim \widetilde{D}_2 \sim D_2 <1$ \cite{Torres2015,Torres2016BJP}. This is similar to what one finds at the metal-insulator transition of noninteracting models~\cite{Ketzmerick1992,Huckestein1994}. 

\begin{figure*}[ht]
\begin{center}
\includegraphics[width=12.cm]{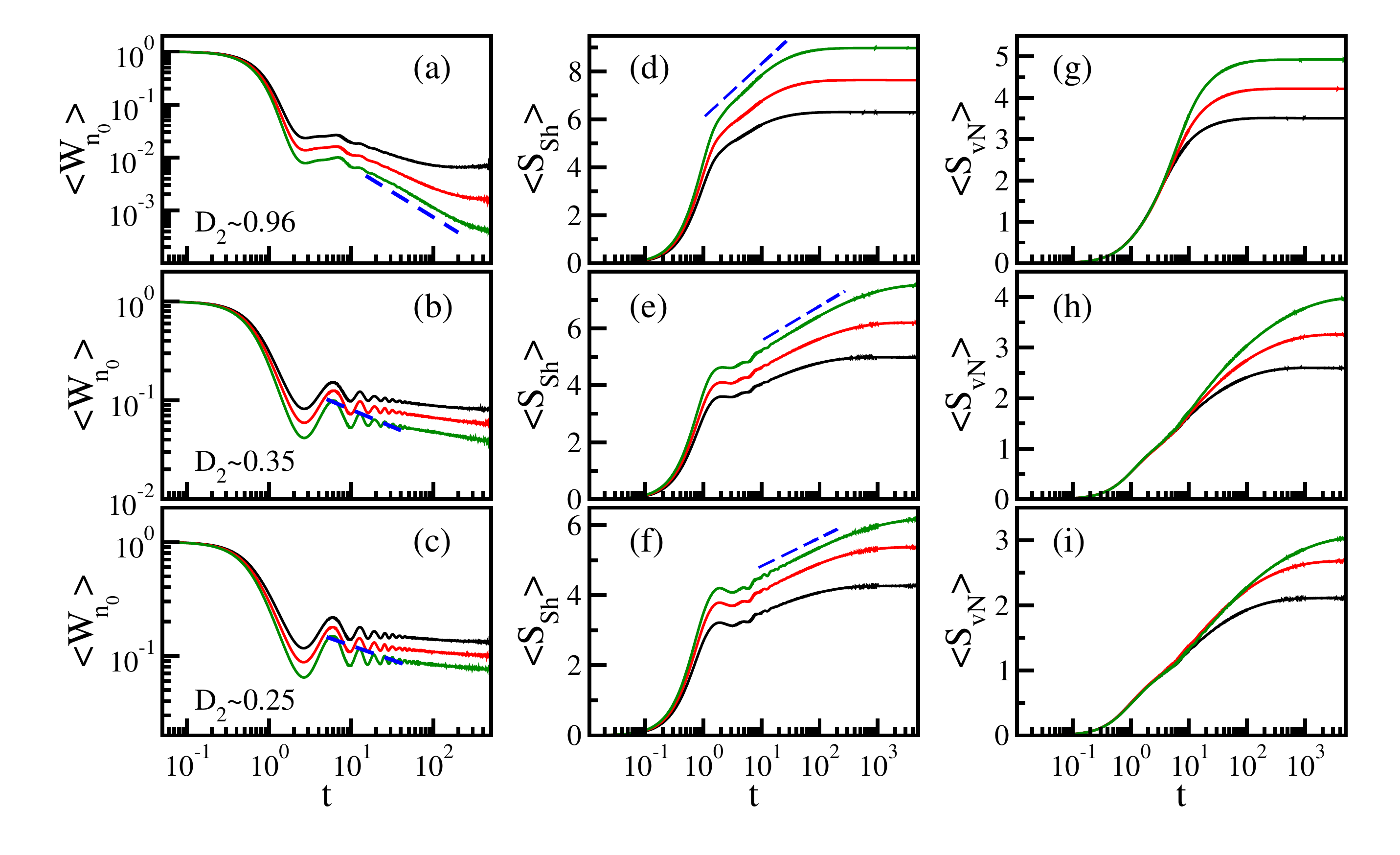}
\vskip -0.7cm
\caption{\label{Fig:SP} Evolution of the average survival probability (a), (b), (c), Shannon entropy in the site-basis (d), (e), (f), and entanglement entropy (g), (h), (i). The strengths of the disorder are: $h=1.25$ in (a), (d), (g); $h=2.5$ in (b), (e), (h); and $h=3.0$ in (c), (f), (i). Curves in (a), (b), (c) from top to bottom: $L=12,14,16$; dashed lines give $\langle W_{n_0}(t) \rangle \propto t^{-\widetilde{D}_2}$. Curves in (d)-(i) from bottom to top: $L=12,14,16$; dashed lines give $\langle S_{Sh}(t) \rangle =A_{Sh} + \widetilde{D}_2 \ln t$, where $A_{Sh}$ is a fitted constant. The averages are done over $10^5$ data of disorder realizations and initial states.}
\end{center}
\end{figure*}

Some illustrations of the behavior of survival probability in region (iii) are provided in Figs.~\ref{Fig:SP} (a), (b), and (c) (more examples can be found in Ref.~\cite{Torres2015}). The initial decay of $\langle W_{n_0} (t) \rangle$ is Gaussian up to $t\sim2$ and then it slows down. For the disorder strengths of the three panels, $\widetilde{D}_2<1$. One sees that for $t>2$, the numerical results for $\langle W_{n_0} (t) \rangle$ (solid lines) agree very well the power-law decay $\propto t^{-\widetilde{D}_2}$ (dashed lines). 

For both entropies, after the quadratic behavior for very short times~\cite{Torres2014NJP,Torres2016}, the evolution becomes linear in $t$ approximately up to where the decay of $\langle W_{n_0}(t) \rangle$ is still Gaussian. The linear increase of the Shannon entropy was studied in \cite{Torres2016,Santos2012PRL,Santos2012PRE,Berman2004,Flambaum2001b}, where semi-analytical expressions were also provided. In region (iii), the linear increase is followed by a logarithmic behavior in time. This occurs for the entanglement entropy [see Figs.~\ref{Fig:SP} (g), (h), (i)], as has been discussed before~\cite{Znidaric2008,Bardarson2012}, but also for the Shannon entropy, as seen in Figs.~\ref{Fig:SP} (d), (e), (f). The analogous behavior reinforces our statement that either one of the entropies can be used in these studies.

Motivated by the power-law decay of the survival probability, $\langle W_{n_0} \rangle \propto t^{-\widetilde{D}_2}$, and by the fact that, from Eq.~(\ref{Eq:Sh_t}), $S_{Sh}(t) = W_{n_0} (t) \ln W_{n_0} (t) + \sum_{n\neq n_0} W_{n} (t) \ln W_{n} (t) $, we fit the logarithmic growth of $\langle S_{Sh}(t) \rangle $ with
\begin{equation}
\langle S_{Sh}(t) \rangle = A_{Sh}  + B_{Sh} \ln (t).
\label{Eq:B}
\end{equation}
Our results indicate a very good agreement between $B_{Sh} $ and $\widetilde{D}_2$. This comparison is done in  Fig.~\ref{Fig:PRini} (b) for $L=14$. We should expect improvement for larger system sizes. In fact, in Figs.~\ref{Fig:SP} (d), (e), (f), the dashed lines correspond to $\langle S_{Sh}(t) \rangle = A_{Sh}  + \widetilde{D}_2 \ln (t)$, with only $A_{Sh}$ being fitted.
Similarly to the power-law decay of $\langle W_{n_0}(t) \rangle$ with $\gamma<1$, the onset of the logarithmic behavior for $\langle S_{Sh}(t) \rangle $ must also be associated with the emergence of correlated eigenstates. This connection deserves further investigation.

\subsection{Time evolution under weak disorder}
\label{Sec:weak}

All the results for region (ii) indicate chaos. The level spacing distribution has the Wigner-Dyson shape, $\langle IPR \rangle$ and $\langle IPR_{n_0} \rangle$ are $\propto N^{-1}$,  and the distribution of the components of the initial state agrees very well with the Porter-Thomas distribution. The quantities that measure the level of chaoticity saturate:  $\langle \eta \rangle$ reaches its smallest values and $D_{1,2}, \widetilde{D}_2 \sim 1$. Yet, the power-law exponent $\gamma$ still varies with $h$. Throughout region (ii), $\gamma > 1$, so it no longer coincides with $\widetilde{D}_2$. As $h$ decreases below 1, $\gamma$ increases monotonically above 1 up to its maximum value $\gamma=2$ \cite{Tavora2016,TavoraARXIV}. 

\begin{figure*}[ht]
\begin{center}
\includegraphics[width=12cm]{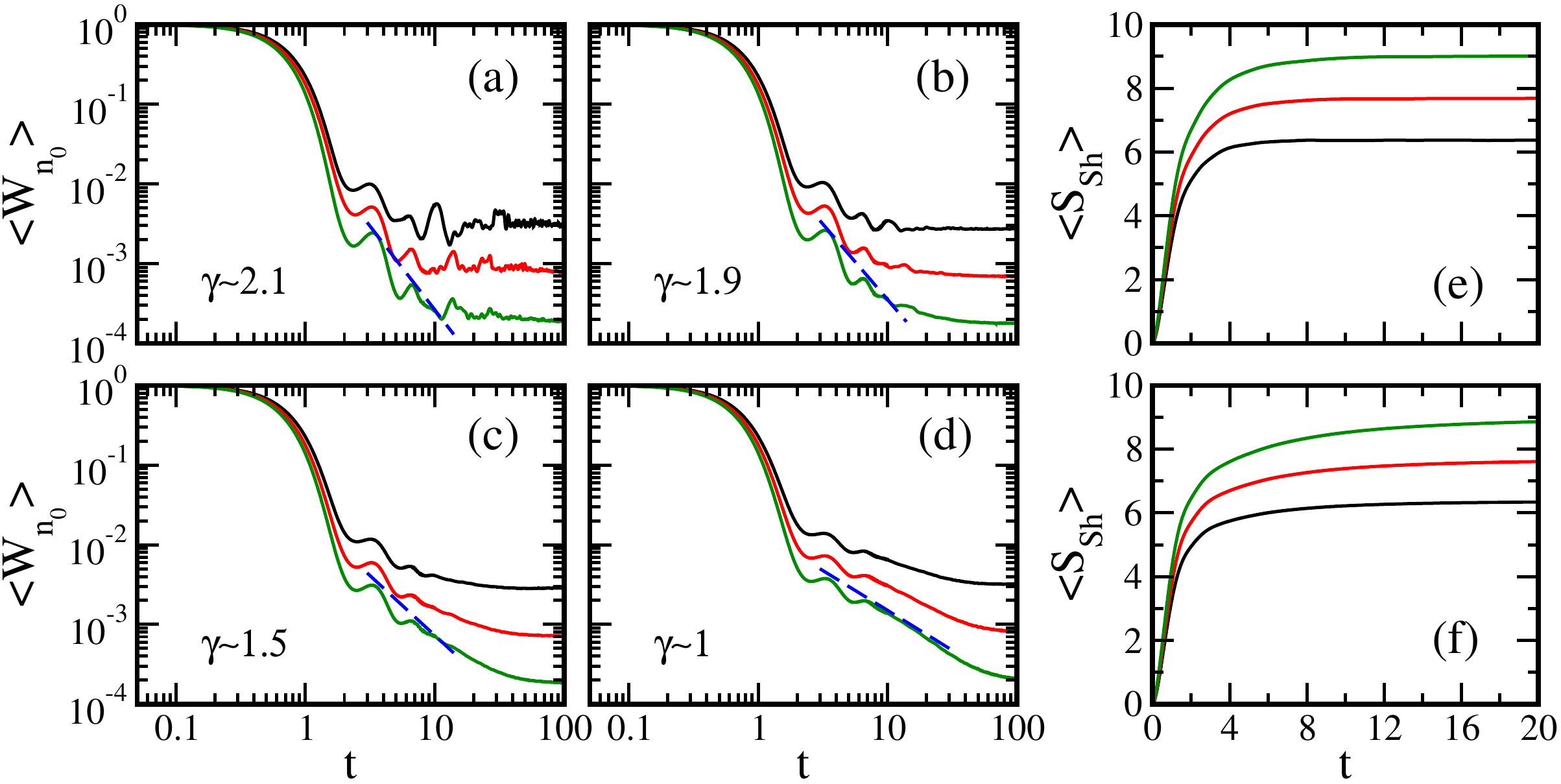}
\caption{\label{Fig:gamma2} 
Evolution of the average survival probability (a), (b), (c), (d) and the Shannon entropy (e), (f). The strengths of the disorder are: $h=0.2$ in (a) and (e); $h=0.4$ in (b); $h=0.6$ in (c); $h=0.8$ in (d) and (f). In (a), (b), (c), (d), the curves from top to bottom are for  $L=12,14,16$. The dashed lines correspond to $\langle W_{n_0} \rangle\propto t^{-\gamma}$ and the value of $\gamma$ is indicated in the panels.  In (e) and (f), the curves from bottom to top are for $L=12,14,16$.  The averages are done over $10^5$ data of disorder realizations and initial states.}
\end{center}
\end{figure*}

The value $\gamma=2$ is a consequence of the Khalfin effect, which refers to the onset of the power-law decay of the survival probability caused by the unavoidable presence of energy bounds in the spectrum~\cite{Khalfin1958}. The phenomenon is well understood in the case of continuous models~\cite{MugaBook,Urbanowski2009,Campo2016}. We have extended these studies to finite many-body quantum systems, where the spectrum is discrete~\cite{Tavora2016,TavoraARXIV}. In this case, when the initial state is chaotic and the LDOS is very well filled, the same sort of analysis used for continuous $\rho_{n_0} (E)$ remains valid. The Fourier transform of a Gaussian LDOS that takes into account the bounds in the spectrum leads asymptotically to $W_{n_0} (t) \propto t^{-2}$. This is the behavior seen in Fig.~\ref{Fig:gamma2} (a). This figure is obtained for $h =0.2$, where $IPR^{FRM}/\langle IPR_{n_0} \rangle$ approaches the largest value for $L=16$. As $h$ increases above $0.2$ up to $1$, $\gamma$ decreases from $2$ and approaches $1$. This progressive reduction of $\gamma$ is shown in Figs.~\ref{Fig:gamma2} (b), (c), and (d).

We expect a chaotic initial state, which implies $\widetilde{D}_2 \sim 1$, to thermalize. A necessary condition for thermalization (that is, for the coincidence between the infinite-time averages of few-body observables and their thermodynamic averages) is the presence of chaotic states, which leads to the unbiased sampling of the Hilbert space~\cite{Borgonovi2016,Alessio2016}. As we saw above, when $\widetilde{D}_2 \sim 1$, the power-law exponent ranges between 1 and 2, while for $\widetilde{D}_2 < 1$, the exponent is smaller than $1$. This indicates that from the value of $\gamma$ we can anticipate whether or not the initial state will thermalize. This point was carefully studied in Refs.~\cite{Tavora2016,TavoraARXIV}.

Combining the results from Sec.~\ref{Sec:strong}  and Sec.~\ref{Sec:weak}, one sees that by varying the strength of the disorder in $H$ (\ref{ham}), all values of $\gamma$ that are reachable by realistic models with two-body interactions, that is $0\leq \gamma \leq 2$, can be covered. These same values can be achieved also with banded random matrices~\cite{Tavora2016,TavoraARXIV}. Examples of these matrices include the power-law banded random matrices that have been widely used in studies of the Anderson metal-insulator transition~\cite{Evers2008} and the Wigner banded random matrices~\cite{Wigner1955}. The variances of their random numbers are large within a bandwidth $b$ around the diagonal elements and very small or zero outside. As $b$ increases from one, all values of $\gamma$ from 0 to 2 are found. In addition, with a very broad $b$ we can go beyond these realistic values of $\gamma$.  The maximum power-law exponent $\gamma =3$  is reached in the limit of full random matrices when $b \rightarrow N$ \cite{Tavora2016,TavoraARXIV,Torres2016}. Banded random matrices provide a general picture of the available power-law exponents for the quench dynamics of isolated many-body quantum systems without being restricted to a particular model. This emphasizes that the results presented here are general and not specific to the spin-1/2 model considered.

In contrast to the survival probability, the entropy growth in time does not provide a clear signature of the Khalfin effect for the system sizes examined here. After the linear increase, where one might expect it to show up, the Shannon entropy and the entanglement entropy simply saturate, as shown for the Shannon entropy in Figs.~\ref{Fig:gamma2} (e) and (f). The survival probability is more sensitive to changes in the spectrum and in the structures of the eigenstates than the entropies.

\subsection{Correlation hole}

The results above show that details about the Hamiltonian and the initial state can be extracted from the dynamics. This is very useful for experiments that routinely study dynamics. The evolution of the survival probability, in particular, offers direct information about level statistics. The presence of level repulsion can be inferred from the difference between the minimum value of the survival probability, $W_{n_0}^{min} $, and its infinite time average, $\overline{W}_{n_0}= IPR_{n_0}$. This difference is known as correlation hole~\cite{Pechukas1984,Leviandier1986,Delon1991,Alhassid1992}. It does not exist in systems with uncorrelated eigenvalues, while in FRM the hole is very deep. The notion of correlation hole was introduced as a method to obtain information about level statistics from systems where one has only partial access to the spectrum~\cite{Pechukas1984,Leviandier1986,Delon1991}.

\begin{figure}[t!]
\includegraphics[width=8.4cm]{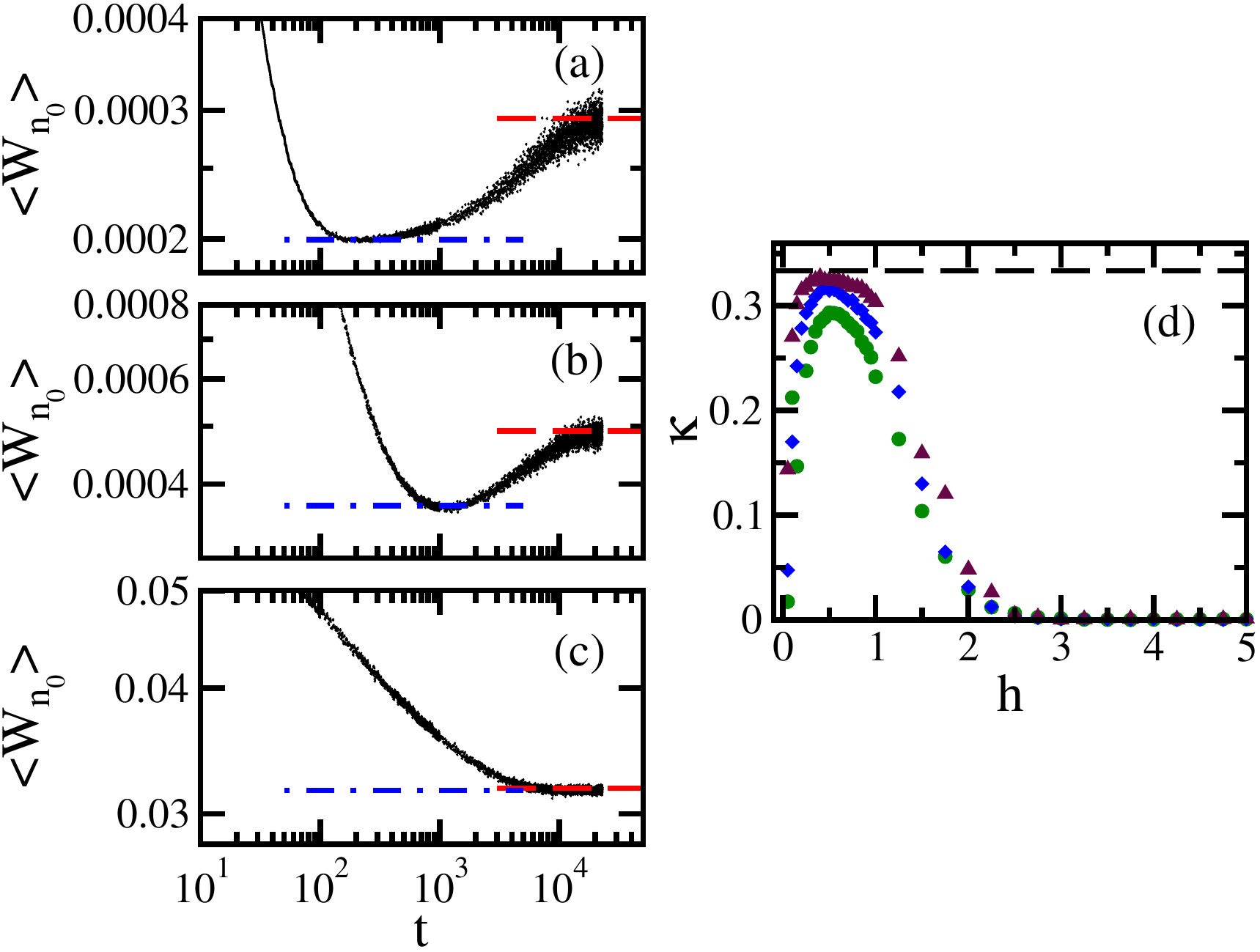}
\caption{\label{Fig:CH}
Average survival probability for long times up to saturation (a,b,c) and the depth $\kappa$ of the correlation hole {\em vs.} disorder strength (d). In (a,b,c) the results are for $L=16$ and $h=0.8,\; 1.25,\;2.50$, respectively. Dashed lines are $\overline{W}_{n_0}=\langle IPR_{n_0}\rangle$ and dot-dashed lines correspond to $\langle W_{n_0}^{min}\rangle$. In (d):  $L=16$ (triangles), $L=14$ (diamonds) and $L=12$ (circles). Dashed line in (d) corresponds to $\kappa^{FRM}=1/3$.}
\end{figure}

In Figs.~\ref{Fig:CH} (a), (b) and (c), we show $\langle W_{n_0}(t) \rangle$ for long times, up to saturation. The disorder strength increases from (a) to (c). The dot-dashed line at the bottom of each panel indicates $\langle W_{n_0}^{min}\rangle $ and the dashed line marks the infinite time average $\overline{W}_{n_0}$. In Fig.~\ref{Fig:CH} (a), where the system is still chaotic ($h=0.8$), the correlation hole is deep. As $h$ increases, the hole fades away and  practically disappears in Fig.~\ref{Fig:CH} (c), where $h=2.5$.

To measure the depth of the correlation hole, we compute
\begin{equation}
\kappa = \frac{\overline{W}_{n_0} - \langle W_{n_0}^{min}\rangle}{\overline{W}_{n_0}} .
\end{equation}
For FRM of GOE, $\overline{W}_{n_0} \sim 3/N$ and $W_{n_0}^{min} \sim 2/N$ \cite{Alhassid1992}, which leads to $\kappa^{FRM} =1/3$.

In Fig.~\ref{Fig:CH} (d), we show the dependence of $\kappa$ on the disorder strength. It peaks in the chaotic region, where it approaches $\kappa^{FRM}$. It decreases in both directions, that of small $h$, where the system gets closer to  integrability, and that of large $h$, where the system approaches localization.

\section{Conclusions}

From the analysis of static and dynamical properties of a finite 1D spin-1/2 system with onsite random disorder, we identified three regions between the integrable limit, where the disorder strength $h$ is zero, and the many-body localized phase. They are: (i) the transition region between the integrable and the chaotic domain, (ii) the chaotic region, and (iii) the intermediate region between chaos and localization. We argued that the eigenstates in region (iii) are delocalized (extended) but nonchaotic (nonergodic). 

Special attention was given to the dynamics, which makes a strong connection with experiments that routinely study dynamics. Information about the spectrum, eigenstates, and initial state can be obtained from the evolution of different quantities. The survival probability, in particular, reveals details that the Shannon information entropy and the von Neumann entanglement entropy cannot capture for the system sizes considered. These include the correlation hole, which is a way to directly detect level repulsion from dynamics, and the Khalfin effect caused by the unavoidable energy bounds of the spectrum.
We also showed that the results for both entropies are comparable, thus either one can be used for the studies of many-body localization. The calculation of the Shannon entropy is more straightforward, because it does not involve any partial trace.

The main characteristics of regions (i), (ii), and (iii) are summarized below.
 
Region (i) has intermediate level statistics (level spacing distribution between Poisson and Wigner-Dyson). This region may disappear in the thermodynamic limit.

Region (ii) shows level repulsion. The eigenstates away from the borders of the spectrum are chaotic, that is $\langle IPR \rangle\propto N^{-1}$. The components of the initial states satisfy the Porter-Thomas distribution and $\langle IPR_{n_0} \rangle\propto N^{-1}$. The power-law decay of the survival probability at long times has exponent $1 \leq \gamma \leq 2$, where $\gamma=2$ indicates ergodic filling of the energy distribution of the initial state. The correlation hole is deep and close to the limits established by random matrix theory. The linear growth of the Shannon entropy and of the von Neumann entanglement entropy is followed by saturation. This is the region where thermalization is expected.

Region (iii) has intermediate level statistics. It seems to become smaller as the system size increases, so it may reduce to a critical point in the thermodynamic limit. In this region, the eigenstates and initial states are delocalized, but nonchaotic, that is $\langle IPR \rangle \propto N^{-D_2}$,  $\langle IPR_{n_0} \rangle \propto N^{-\widetilde{D}_2}$ with $D_2 \sim \widetilde{D}_2 <1$. From the analysis of the entropies, we also find that $D_1<1$. Fractal dimensions smaller than $1$ indicate that the states are fractal and therefore nonergodic. This lack of ergodicity is reflected into the dynamics. The power-law exponent of the long-time decay of $W_{n_0}(t)$ is smaller than $1$ and agrees with the fractal dimension, $\gamma \sim \widetilde{D}_2$. The linear increase in time of the entropies is followed by a logarithmic growth before saturation. Both $\widetilde{D}_2<1$ and the logarithmic behavior are caused by correlations in the eigenstates. This is the region where one expects subdiffusive transport.

\bigskip
\section*{Acknowledgments}
EJTH acknowledges funding from CONACyT, PRODEP-SEP and Proyectos VIEP-BUAP 2016, Mexico. EJTH is also grateful to LNS-BUAP for allowing use of their supercomputing facility. This work was supported by the NSF grants No. DMR-1147430 and No. DMR-1603418.

\bibliographystyle{andp2012}

\begin{thebibliography}{[100]}

\bibitem{Anderson1958}
 \textsc{P.\,W. Anderson} \jr{Phys. Rev.} \textbf{109}, 1492 (1958).


\bibitem{Lee1985}
 \textsc{P.\,A. Lee} and  \textsc{T.\,V. Ramakhrishnan} \jr{Rev. Mod. Phys.}
  \textbf{57}, 287 (1985).


\bibitem{Kramer1993}
 \textsc{B.~Kramer} and  \textsc{A.~MacKinnon} \jr{Rep. Prog. Phys.}
  \textbf{56}, 1469 (1993).


\bibitem{Billy2008}
 \textsc{J.~Billy},  \textsc{V.~Josse},  \textsc{Z.~Zuo},  \textsc{A.~Bernard},
   \textsc{B.~Hambrecht},  \textsc{P.~Lugan},  \textsc{D.~Clement},
  \textsc{L.~Sanchez-Palencia},  \textsc{P.~Bouyer},  and  \textsc{A.~Aspect}
  \jr{Nature} \textbf{453}, 891 (2008).


\bibitem{Roati2008}
 \textsc{G.~Roati},  \textsc{C.~D'Errico},  \textsc{L.~Fallani},
  \textsc{M.~Fattori},  \textsc{C.~Fort},  \textsc{M.~Zaccanti},
  \textsc{G.~Modugno},  \textsc{M.~Modugno},  and  \textsc{M.~Inguscio}
  \jr{Nature} \textbf{453}, 895 (2008).


\bibitem{Modugno2010}
 \textsc{G.~Modugno} \jr{Rep. Progr. Phys.} \textbf{73}, 102401 (2010).


\bibitem{McGehee2013}
 \textsc{W.\,R. McGehee},  \textsc{S.\,S. Kondov},  \textsc{W.~Xu},
  \textsc{J.\,J. Zirbel},  and  \textsc{B.~DeMarco} \jr{Phys. Rev. Lett.}
  \textbf{111}, 145303 (2013).


\bibitem{Fleishman1980}
 \textsc{L.~Fleishman} and  \textsc{P.\,W. Anderson} \jr{Phys. Rev. B}
  \textbf{21}, 2366 (1980).


\bibitem{Altshuler1988}
 \textsc{B.\,L. Altshuler},  \textsc{I.\,K. Zharakeshev},  \textsc{S.\,A.
  Kotochigova},  and  \textsc{B.\,I. Shklovskii} \jr{Sov. Phys. JETP}
  \textbf{67}, 625 (1988).


\bibitem{Giamarchi1988}
 \textsc{T.~Giamarchi} and  \textsc{H.\,J. Schulz} \jr{Phys. Rev. B}
  \textbf{37}, 325(1988).


\bibitem{Dorokhov1990}
 \textsc{O.\,N. Dorokhov} \jr{Sov. Phys. JETP} \textbf{71}, 360 (1990).


\bibitem{Shepelyansky1994}
 \textsc{D.\,L. Shepelyansky} \jr{Phys. Rev. Lett.} \textbf{73}, 2607 (1994).


\bibitem{Imry1995}
 \textsc{Y.~Imry} \jr{Europhys. Lett.} \textbf{30}, 405 (1995).


\bibitem{Altshuler1997}
 \textsc{B.\,L. Altshuler},  \textsc{Y.~Gefen},  \textsc{A.~Kamenev},  and
  \textsc{L.\,S. Levitov} \jr{Phys. Rev. Lett.} \textbf{78}, 2803(1997).


\bibitem{Guhr1998}
 \textsc{T.~Guhr},  \textsc{A.~Mueller-Gr\"oeling},  and  \textsc{H.\,A.
  Weidenm\"uller} \jr{Phys. Rep.} \textbf{299}, 189 (1998).


\bibitem{Georgeot2000}
 \textsc{B.~Georgeot} and  \textsc{D.\,L. Shepelyansky} \jr{Phys. Rev. E}
  \textbf{62}, 6366 (2000).


\bibitem{Berman2001}
 \textsc{G.\,P. Berman},  \textsc{F.~Borgonovi},  \textsc{F.\,M. Izrailev},
  and  \textsc{V.\,I. Tsifrinovich} \jr{Phys. Rev. E.} \textbf{64}, 056226
  (2001).


\bibitem{Santos2004}
 \textsc{L.\,F. Santos} \jr{J. Phys. A} \textbf{37}, 4723 (2004).


\bibitem{SantosEscobar2004}
 \textsc{L.\,F. Santos},  \textsc{G.~Rigolin},  and  \textsc{C.\,O. Escobar}
  \jr{Phys. Rev. A} \textbf{69}, 042304 (2004).


\bibitem{Santos2005loc}
 \textsc{L.\,F. Santos},  \textsc{M.\,I. Dykman},  \textsc{M.~Shapiro},  and
  \textsc{F.\,M. Izrailev} \jr{Phys. Rev. A} \textbf{71}, 012317 (2005).


\bibitem{Gornyi2005}
 \textsc{I.\,V. Gornyi},  \textsc{A.\,D. Mirlin},  and  \textsc{D.\,G.
  Polyakov} \jr{Phys. Rev. Lett.} \textbf{95}, 206603 (2005).


\bibitem{Basko2006}
 \textsc{D.\,M. Basko},  \textsc{I.\,L. Aleiner},  and  \textsc{B.\,L.
  Altshuler} \jr{Ann. Phys.} \textbf{321}, 1126 (2006).


\bibitem{Oganesyan2007}
 \textsc{V.~Oganesyan} and  \textsc{D.\,A. Huse} \jr{Phys. Rev. B} \textbf{75},
  155111 (2007).


\bibitem{Znidaric2008}
 \textsc{M.~\ifmmode\,\check{Z}\else\,\v{Z}\fi{}nidari\ifmmode\,\check{c}\else
  \v{c}\fi{}},  \textsc{T.~Prosen},  and
  \textsc{P.~Prelov\ifmmode\,\check{s}\else \v{s}\fi{}ek} \jr{Phys. Rev. B}
  \textbf{77}, 064426 (2008).


\bibitem{Brown2008}
 \textsc{W.\,G. Brown},  \textsc{L.\,F. Santos},  \textsc{D.~Starling},  and
  \textsc{L.~Viola} \jr{Phys. Rev. E} \textbf{77}, 021106 (2008).


\bibitem{Dukesz2009}
 \textsc{F.~Dukesz},  \textsc{M.~Zilbergerts},  and  \textsc{L.\,F. Santos}
  \jr{New J. Phys.} \textbf{11}, 043026 (2009).


\bibitem{Pal2010}
 \textsc{A.~Pal} and  \textsc{D.\,A. Huse} \jr{Phys. Rev. B} \textbf{82},
  174411 (2010).


\bibitem{Bardarson2012}
 \textsc{J.\,H. Bardarson},  \textsc{F.~Pollmann},  and  \textsc{J.\,E. Moore}
  \jr{Phys. Rev. Lett.} \textbf{109}, 017202 (2012).


\bibitem{DeLuca2013}
 \textsc{A.\,D. Luca} and  \textsc{A.~Scardicchio} \jr{Europhys. Lett.}
  \textbf{101}, 37003 (2013).


\bibitem{Imbrie2016a}
 \textsc{J.\,Z. Imbrie} \jr{J. Stat. Phys.} \textbf{163}, 998 (2016).


\bibitem{Imbrie2016b}
 \textsc{J.\,Z. Imbrie} \jr{Phys. Rev. Lett.} \textbf{117}, 027201 (2016).


\bibitem{Huse2014}
 \textsc{D.\,A. Huse},  \textsc{R.~Nandkishore},  and  \textsc{V.~Oganesyan}
  \jr{Phys. Rev. B} \textbf{90}, 174202 (2014).


\bibitem{Kjall2014}
 \textsc{J.\,A. Kj\"all},  \textsc{J.\,H. Bardarson},  and
  \textsc{F.~Pollmann} \jr{Phys. Rev. Lett.} \textbf{113}, 107204 (2014).


\bibitem{Lev2014}
 \textsc{Y.~Bar~Lev} and  \textsc{D.\,R. Reichman} \jr{Phys. Rev. B}
  \textbf{89}, 220201 (2014).
  
  
\bibitem{Lev2015}
 \textsc{Y.~Bar~Lev}, \textsc{G. Cohen},  and  \textsc{D.\,R. Reichman} \jr{Phys. Rev. Lett.}
  \textbf{114}, 100601 (2015).  


\othercit
\bibitem{GroverARXIV}
 \textsc{T.~Grover},
Certain general constraints on the many-body localization transition,
arXiv:1405.1471.


\bibitem{Vosk2015}
 \textsc{R.~Vosk},  \textsc{D.\,A. Huse},  and  \textsc{E.~Altman} \jr{Phys.
  Rev. X} \textbf{5}, 031032 (2015).


\bibitem{Chandran2015}
 \textsc{A.~Chandran} and  \textsc{C.\,R. Laumann} \jr{Phys. Rev. B}
  \textbf{92}, 024301 (2015).


\bibitem{Luitz2015}
 \textsc{D.\,J. Luitz},  \textsc{N.~Laflorencie},  and  \textsc{F.~Alet}
  \jr{Phys. Rev. B} \textbf{91}, 081103 (2015).


\bibitem{Luitz2016}
 \textsc{D.\,J. Luitz},  \textsc{N.~Laflorencie},  and  \textsc{F.~Alet}
  \jr{Phys. Rev. B} \textbf{93}, 060201 (2016).


\bibitem{Ros2015}
 \textsc{V.~Ros},  \textsc{M.~M\"uller},  and  \textsc{A.~Scardicchio}
  \jr{Nucl. Phys. B} \textbf{891}, 420 (2015).


\bibitem{Torres2015}
 \textsc{E.\,J. Torres-Herrera} and  \textsc{L.\,F. Santos} \jr{Phys. Rev. B}
  \textbf{92}, 014208 (2015).


\bibitem{Torres2016BJP}
 \textsc{E.\,J. Torres-Herrera},  \textsc{M.~T\'avora},  and  \textsc{L.\,F.
  Santos} \jr{Braz. J. Phys.} \textbf{46}, 239 (2016).


\bibitem{Yang2015}
 \textsc{Z.\,C. Yang},  \textsc{C.~Chamon},  \textsc{A.~Hamma},  and
  \textsc{E.\,R. Mucciolo} \jr{Phys. Rev. Lett.} \textbf{115}, 267206
  (2015).


\bibitem{Altman2015}
 \textsc{E.~Altman} and  \textsc{R.~Vosk} \jr{Ann. Rev. Cond. Mat. Phys.}
  \textbf{6}, 383 (2015).


\bibitem{Nandkishore2015}
 \textsc{R.~Nandkishore} and  \textsc{D.~Huse} \jr{Annu. Rev. Condens. Matter
  Phys.} \textbf{6}, 15 (2015).


\bibitem{Serbyn2013}
 \textsc{M.~Serbyn},  \textsc{Z.~Papi\ifmmode\,\acute{c}\else \'{c}\fi{}},  and
   \textsc{D.\,A. Abanin} \jr{Phys. Rev. Lett.} \textbf{111}, 127201 (2013).


\bibitem{Serbyn2014}
 \textsc{M.~Serbyn},  \textsc{Z.~Papi\ifmmode\,\acute{c}\else \'{c}\fi{}},  and
   \textsc{D.\,A. Abanin} \jr{Phys. Rev. B} \textbf{90}, 174302 (2014).


\bibitem{Serbyn2015}
 \textsc{M.~Serbyn},  \textsc{Z.~Papi\ifmmode\,\acute{c}\else \'{c}\fi{}},  and
   \textsc{D.\,A. Abanin} \jr{Phys. Rev. X} \textbf{5}, 041047 (2015).


\bibitem{Serbyn2016}
 \textsc{M.~Serbyn} and  \textsc{J.\,E. Moore} \jr{Phys. Rev. B} \textbf{93},
  041424 (2016).


\bibitem{Singh2016}
 \textsc{R.~Singh},  \textsc{J.\,H. Bardarson},  and  \textsc{F.~Pollmann}
  \jr{New J. Phys.} \textbf{18}, 023046 (2016).


\bibitem{Agarwal2015}
 \textsc{K.~Agarwal},  \textsc{S.~Gopalakrishnan},  \textsc{M.~Knap},
  \textsc{M.~M\"uller},  and  \textsc{E.~Demler} \jr{Phys. Rev. Lett.}
  \textbf{114}, 160401 (2015).


\bibitem{Gopalakrishnan2016}
 \textsc{S.~Gopalakrishnan},  \textsc{K.~Agarwal},  \textsc{E.\,A. Demler},
  \textsc{D.\,A. Huse},  and  \textsc{M.~Knap} \jr{Phys. Rev. B} \textbf{93},
  134206 (2016).


\bibitem{Monthus2016}
 \textsc{C.~Monthus} \jr{Entropy} \textbf{18}, 122 (2016).


\bibitem{Xiaopeng2015}
 \textsc{X.~Li},  \textsc{S.~Ganeshan},  \textsc{J.\,H. Pixley},  and
  \textsc{S.~Das~Sarma} \jr{Phys. Rev. Lett.} \textbf{115}, 186601 (2015).


\bibitem{Xiaopeng2016}
 \textsc{X.~Li},  \textsc{J.\,H. Pixley},  \textsc{D.\,L. Deng},
  \textsc{S.~Ganeshan},  and  \textsc{S.~Das~Sarma} \jr{Phys. Rev. B}
  \textbf{93}, 184204 (2016).


\bibitem{Devakul2015}
 \textsc{T.~Devakul} and  \textsc{R.\,R.\,P. Singh} \jr{Phys. Rev. Lett.}
  \textbf{115}, 187201 (2015).


\bibitem{Goold2015}
 \textsc{J.~Goold},  \textsc{C.~Gogolin},  \textsc{S.\,R. Clark},
  \textsc{J.~Eisert},  \textsc{A.~Scardicchio},  and  \textsc{A.~Silva}
  \jr{Phys. Rev. B} \textbf{92}, 180202 (2015).


\bibitem{Santos2016}
 \textsc{L.\,F. Santos},  \textsc{F.~Borgonovi},  and  \textsc{G.\,L. Celardo}
  \jr{Phys. Rev. Lett.} \textbf{116}, 250402 (2016).


\bibitem{Bertrand2016}
 \textsc{C.\,L. Bertrand} and  \textsc{A.\,M. Garc\'{\i}a-Garc\'{\i}a}
  \jr{Phys. Rev. B} \textbf{94}, 144201 (2016).


\bibitem{Gogolin2016}
 \textsc{C.~Gogolin} and  \textsc{J.~Eisert} \jr{Rep. Prog. Phys.}
  \textbf{79}, 056001 (2016).


\bibitem{Cohen2016}
 \textsc{D.~Cohen},  \textsc{V.\,I. Yukalov},  and  \textsc{K.~Ziegler}
  \jr{Phys. Rev. A} \textbf{93}, 042101 (2016).


\bibitem{Barisic2016}
 \textsc{O.\,S.
  Bari\ifmmode\,\check{s}\else\,\v{s}\fi{}i\ifmmode\,\acute{c}\else
  \'{c}\fi{}},  \textsc{J.~Kokalj},  \textsc{I.~Balog},  and
  \textsc{P.~Prelov\ifmmode\,\check{s}\else \v{s}\fi{}ek} \jr{Phys. Rev. B}
  \textbf{94}, 045126 (2016).


\bibitem{Khemani2016}
 \textsc{V.~Khemani},  \textsc{F.~Pollmann},  and  \textsc{S.\,L. Sondhi}
  \jr{Phys. Rev. Lett.} \textbf{116}, 247204 (2016).


\othercit
\bibitem{EnssARXIV}
 \textsc{T.~Enss},  \textsc{F.~Andraschko},  and  \textsc{J.~Sirker},
Many-body localization in infinite chains,
arXiv:1608.05733.


\bibitem{Schreiber2015}
 \textsc{M.~Schreiber},  \textsc{S.\,S. Hodgman},  \textsc{P.~Bordia},
  \textsc{H.\,P. L\"uschen},  \textsc{M.\,H. Fischer},  \textsc{R.~Vosk},
  \textsc{E.~Altman},  \textsc{U.~Schneider},  and  \textsc{I.~Bloch}
  \jr{Science} \textbf{349}, 842 (2015).


\bibitem{Kondov2015}
 \textsc{S.\,S. Kondov},  \textsc{W.\,R. McGehee},  \textsc{W.~Xu},  and
  \textsc{B.~DeMarco} \jr{Phys. Rev. Lett.} \textbf{114}, 083002 (2015).


\bibitem{Bordia2016}
 \textsc{P.~Bordia},  \textsc{H.\,P. L\"uschen},  \textsc{S.\,S. Hodgman},
  \textsc{M.~Schreiber},  \textsc{I.~Bloch},  and  \textsc{U.~Schneider}
  \jr{Phys. Rev. Lett.} \textbf{116}, 140401 (2016).


\bibitem{Smith2015}
 \textsc{J.~Smith},  \textsc{A.~Lee},  \textsc{P.~Richerme},
  \textsc{B.~Neyenhuis},  \textsc{P.\,W. Hess},  \textsc{P.~Hauke},
  \textsc{M.~Heyl},  \textsc{D.\,A. Huse},  and  \textsc{C.~Monroe} \jr{Nat.
  Phys.} \textbf{12}, 907 (2016).


\bibitem{DeLuca2014}
 \textsc{A.~De~Luca},  \textsc{B.\,L. Altshuler},  \textsc{V.\,E. Kravtsov},
  and  \textsc{A.~Scardicchio} \jr{Phys. Rev. Lett.} \textbf{113}, 046806
  (2014).


\bibitem{Kravtsov2015}
 \textsc{V.\,E. Kravtsov},  \textsc{I.\,M. Khaymovich},  \textsc{E.~Cuevas},
  and  \textsc{M.~Amini} \jr{New J. Phys.} \textbf{17}, 122002 (2015).


\bibitem{Znidaric2016}
 \textsc{M.~\ifmmode\,\check{Z}\else\,\v{Z}\fi{}nidari\ifmmode\,\check{c}\else
  \v{c}\fi{}},  \textsc{A.~Scardicchio},  and  \textsc{V.\,K. Varma} \jr{Phys.
  Rev. Lett.} \textbf{117}, 040601 (2016).


\bibitem{Wigner1958}
 \textsc{E.\,P. Wigner} \jr{Ann. Math.} \textbf{67}, 325 (1958).


\othercit
\bibitem{MehtaBook}
 \textsc{M.\,L. Mehta},
Random Matrices (Academic Press, Boston, 1991).


\bibitem{ZelevinskyRep1996}
 \textsc{V.~Zelevinsky},  \textsc{B.\,A. Brown},  \textsc{N.~Frazier},  and
  \textsc{M.~Horoi} \jr{Phys. Rep.} \textbf{276}, 85 (1996).
  
\bibitem{Torres2016}
 \textsc{E.\,J. Torres-Herrera},  \textsc{J.~Karp},  \textsc{M.~T\'avora},  and
   \textsc{L.\,F. Santos} \jr{Entropy} \textbf{18}(10), 359 (2016).  


\bibitem{Brody1981}
 \textsc{T.\,A. Brody},  \textsc{J.~Flores},  \textsc{J.\,B. French},
  \textsc{P.\,A. Mello},  \textsc{A.~Pandey},  and  \textsc{S.\,S.\,M. Wong}
  \jr{Rev. Mod. Phys} \textbf{53}, 385 (1981).


\bibitem{Avishai2002}
 \textsc{Y.~Avishai},  \textsc{J.~Richert},  and  \textsc{R.~Berkovitz}
  \jr{Phys. Rev. B} \textbf{66}, 052416 (2002).


\bibitem{Tavora2016}
 \textsc{M.~T\'avora},  \textsc{E.\,J. Torres-Herrera},  and  \textsc{L.\,F.
  Santos} \jr{Phys. Rev. A} \textbf{94}, 041603R (2016).


\othercit
\bibitem{TavoraARXIV}
 \textsc{M.~T\'avora},  \textsc{E.\,J. Torres-Herrera},  and  \textsc{L.\,F.
  Santos},
Power-law decay exponents: a dynamical criterion for predicting thermalization,
arXiv:1610.04240.


\bibitem{Pechukas1984}
 \textsc{P.~Pechukas} \jr{J. Phys. Chem.} \textbf{88}, 4823 (1984).


\bibitem{Leviandier1986}
 \textsc{L.~Leviandier},  \textsc{M.~Lombardi},  \textsc{R.~Jost},  and
  \textsc{J.\,P. Pique} \jr{Phys. Rev. Lett.} \textbf{56}, 2449
  (1986).


\bibitem{Delon1991}
 \textsc{A.~Delon},  \textsc{R.~Jost},  and  \textsc{M.~Lombardi} \jr{J. Chem.
  Phys.} \textbf{95}, 5701 (1991).


\bibitem{Alhassid1992}
 \textsc{Y.~Alhassid} and  \textsc{R.\,D. Levine} \jr{Phys. Rev. A}
  \textbf{46}, 4650 (1992).


\bibitem{Izrailev1989}
 \textsc{F.\,M. Izrailev} \jr{J. Phys. A} \textbf{22}, 865 (1989).


\bibitem{Shklovskii1993}
 \textsc{B.\,I. Shklovskii},  \textsc{B.~Shapiro},  \textsc{B.\,R. Sears},
  \textsc{P.~Lambrianides},  and  \textsc{H.\,B. Shore} \jr{Phys. Rev. B}
  \textbf{47}, 11487 (1993).


\bibitem{Gubin2012}
 \textsc{A.~Gubin} and  \textsc{L.\,F. Santos} \jr{Am. J. Phys.} \textbf{80},
  246 (2012).


\bibitem{Santos2009JMP}
 \textsc{L.\,F. Santos} \jr{J. Math. Phys} \textbf{50}, 095211 (2009).


\bibitem{Zangara2013}
 \textsc{P.\,R. Zangara},  \textsc{A.\,D. Dente},  \textsc{E.\,J.
  Torres-Herrera},  \textsc{H.\,M. Pastawski},  \textsc{A.~Iucci},  and
  \textsc{L.\,F. Santos} \jr{Phys. Rev. E} \textbf{88}, 032913 (2013).


\bibitem{Jacquod1997}
 \textsc{P.~Jacquod} and  \textsc{D.\,L. Shepelyansky} \jr{Phys. Rev. Lett.}
  \textbf{79}, 1837 (1997).


\bibitem{Santos2010PRE}
 \textsc{L.\,F. Santos} and  \textsc{M.~Rigol} \jr{Phys. Rev. E} \textbf{81},
  036206 (2010).


\bibitem{Torres2014PRE}
 \textsc{E.\,J. Torres-Herrera} and  \textsc{L.\,F. Santos} \jr{Phys. Rev. E}
  \textbf{89}, 062110 (2014).


\bibitem{Gomez2002}
 \textsc{J.\,M.\,G. G\'omez},  \textsc{R.\,A. Molina},  \textsc{A.~Rela\~no},
  and  \textsc{J.~Retamosa} \jr{Phys. Rev. E} \textbf{66}, 036209 (2002).


\bibitem{Evers2008}
 \textsc{F.~Evers} and  \textsc{A.\,D. Mirlin} \jr{Rev. Mod. Phys.}
  \textbf{80}, 1355 (2008).


\bibitem{Atas2014}
 \textsc{Y.\,Y. Atas} and  \textsc{E.~Bogomolny} \jr{Phil. Trans. R. Soc. A}
  \textbf{372} (2014).


\othercit
\bibitem{footDelta}
For fixed $h$, $IPR^{FRM}/\langle IPR \rangle$ decreases monotonically as the
  ratio between the Ising interaction strength and the flip-flop term strength
  increases~\cite{Dukesz2009}.


\bibitem{Torres2013}
 \textsc{E.\,J. Torres-Herrera} and  \textsc{L.\,F. Santos} \jr{Phys. Rev. E}
  \textbf{88}, 042121 (2013).


\bibitem{Santos2012PRL}
 \textsc{L.\,F. Santos},  \textsc{F.~Borgonovi},  and  \textsc{F.\,M. Izrailev}
  \jr{Phys. Rev. Lett.} \textbf{108}, 094102 (2012).


\bibitem{Santos2012PRE}
 \textsc{L.\,F. Santos},  \textsc{F.~Borgonovi},  and  \textsc{F.\,M. Izrailev}
  \jr{Phys. Rev. E} \textbf{85}, 036209 (2012).


\bibitem{He2013}
 \textsc{K.~He} and  \textsc{M.~Rigol} \jr{Phys. Rev. A} \textbf{87}, 043615
  (2013).


\bibitem{Rigol2014}
 \textsc{M.~Rigol} \jr{Phys. Rev. Lett.} \textbf{112}, 170601 (2014).


\bibitem{Amico2008}
 \textsc{L.~Amico},  \textsc{R.~Fazio},  \textsc{A.~Osterloh},  and
  \textsc{V.~Vedral} \jr{Rev. Mod. Phys.} \textbf{80}, 517 (2008).


\bibitem{Laflorencie2016}
 \textsc{N.~Laflorencie} \jr{Phys. Rep.} \textbf{646}, 1  (2016).


\bibitem{Page1993}
 \textsc{D.\,N. Page} \jr{Phys. Rev. Lett.} \textbf{71}, 1291
  (1993).


\othercit
\bibitem{footSd}
The study in~\cite{DeLuca2013} actually considered the diagonal entropy, which
  is the Shannon entropy of the initial state written in the energy
  eigenbasis~\cite{Polkovnikov2011,Santos2011PRL}.


\bibitem{Torres2014PRA}
 \textsc{E.\,J. Torres-Herrera} and  \textsc{L.\,F. Santos} \jr{Phys. Rev. A}
  \textbf{89}, 043620 (2014).


\bibitem{Torres2014NJP}
 \textsc{E.\,J. Torres-Herrera},  \textsc{M.~Vyas},  and  \textsc{L.\,F.
  Santos} \jr{New J. Phys.} \textbf{16}, 063010 (2014).


\bibitem{Torres2014PRAb}
 \textsc{E.\,J. Torres-Herrera} and  \textsc{L.\,F. Santos} \jr{Phys. Rev. A}
  \textbf{90}, 033623 (2014).


\othercit
\bibitem{PorterBook}
 \textsc{C.\,E. Porter},
Statistical Theories of Spectra: Fluctuations (Academic Press, New York, 1965).


\bibitem{Ketzmerick1992}
 \textsc{R.~Ketzmerick},  \textsc{G.~Petschel},  and  \textsc{T.~Geisel}
  \jr{Phys. Rev. Lett.} \textbf{69}, 695 (1992).


\bibitem{Huckestein1994}
 \textsc{B.~Huckestein} and  \textsc{L.~Schweitzer} \jr{Phys. Rev. Lett.}
  \textbf{72}, 713 (1994).


\bibitem{Berman2004}
 \textsc{G.\,P. Berman},  \textsc{F.~Borgonovi},  \textsc{F.\,M. Izrailev},
  and  \textsc{A.~Smerzi} \jr{Phys. Rev. Lett.} \textbf{92}, 030404
  (2004).


\bibitem{Flambaum2001b}
 \textsc{V.\,V. Flambaum} and  \textsc{F.\,M. Izrailev} \jr{Phys. Rev. E}
  \textbf{64}, 036220 (2001).


\bibitem{Khalfin1958}
 \textsc{L.\,A. Khalfin} \jr{Sov. Phys. JETP} \textbf{6}, 1053 (1958).


\othercit
\bibitem{MugaBook}
 \textsc{J.\,G. Muga},  \textsc{A.~Ruschhaupt},  and  \textsc{A.~del Campo},
Time in Quantum Mechanics, vol. 2 (Springer, London, 2009).


\bibitem{Urbanowski2009}
 \textsc{K.~Urbanowski} \jr{Eur. Phys. J. D} \textbf{54}, 25 (2009).


\bibitem{Campo2016}
 \textsc{A.~del Campo} \jr{New J. Phy.} \textbf{18}, 015014 (2016).


\bibitem{Borgonovi2016}
 \textsc{F.~Borgonovi},  \textsc{F.\,M. Izrailev},  \textsc{L.\,F. Santos},
  and  \textsc{V.\,G. Zelevinsky} \jr{Phys. Rep.} \textbf{626}, 1 (2016).


\bibitem{Alessio2016}
 \textsc{L.~D'Alessio},  \textsc{Y.~Kafri},  \textsc{A.~Polkovnikov},  and
  \textsc{M.~Rigol} \jr{Adv. Phys.} \textbf{65}, 239 (2016).


\bibitem{Wigner1955}
 \textsc{E.\,P. Wigner} \jr{Ann. Math.} \textbf{62}, 548 (1955).


\bibitem{Polkovnikov2011}
 \textsc{A.~Polkovnikov} \jr{Ann. Phys. (N.Y.)} \textbf{326}, 486 (2011).


\bibitem{Santos2011PRL}
 \textsc{L.\,F. Santos},  \textsc{A.~Polkovnikov},  and  \textsc{M.~Rigol}
  \jr{Phys. Rev. Lett.} \textbf{107}, 040601 (2011).

\end{thebibliography}

\providecommand{\WileyBibTextsc}{}
\let\textsc\WileyBibTextsc
\providecommand{\othercit}{}
\providecommand{\jr}[1]{#1}
\providecommand{\etal}{~et~al.}

\end{document}